\documentclass{aa}

\usepackage{graphicx}
\def \aap{A\& A}
\def \aj{AJ}
\def \apj{ApJ}
\def \apjl{ApJ}
\def \apjs{ApJS}
\def \araa{ARA\& A}
\def \mnras{MNRAS}
\def \nat{Nature}
\def \prd{Phys. Rev. D}

\begin{document}
\title{Numerical simulations of the cosmic star formation history}
\titlerunning{Cosmic star formation in simulations}
\author{Y. Ascasibar\inst{1,2} \and G. Yepes\inst{1} \and S.
 Gottl{\"o}ber\inst{2} \and V. M{\"u}ller\inst{2}}
\institute{ Grupo de Astrof{\'\i}sica, C-XI Universidad Aut{\'o}noma de Madrid,
 E--28049 Spain \and Astrophysikalisches Institut Potsdam, 
 An der Sternwarte 16, D--14482 Germany}
\offprints{Y. Ascasibar\\ \email{yago.ascasibar@uam.es}}
\date{Received 12 November 2001 , accepted 27 February 2002}

\abstract{
The cosmic star formation history in Cold Dark Matter dominated cosmological
scenarios is studied by means of hydrodynamical numerical simulations. In 
particular, we explore a low density model with a $\Lambda$-term and two high 
density models with different power spectra, all of them being spatially flat. 
Our simulations employ a fully nonlinear N-body and Eulerian hydrodynamics 
algorithm with a model for star formation and supernovae feedback that depends 
on two phenomenological parameters determined in agreement with previous 
papers. We find a nearly constant star formation rate beyond $z=1$, and we 
discuss which facts may determine the decrease in the SFR from $z=1$ to the 
present epoch. The $\Lambda$-term cosmology with realistic parameters for star 
formation and feedback best reproduces the observed star formation history. 
}
\maketitle

\section{Introduction}

A detailed understanding of galaxy formation remains one of the primary goals of
modern cosmology.  Whereas gravitation is basically responsible for the large
scale structure of the universe, many physical phenomena, covering a broad
dynamical range, are involved on galactic scales.  A reliable treatment of
baryonic physics is essential in order to obtain a consistent picture of galaxy
formation and evolution, where star formation and feedback to the interstellar
medium (ISM) play a key role.

The star formation history of the universe, i.e.  the global star formation rate
(SFR) density as a function of redshift, is a crucial test for galaxy formation
scenarios.  Many measurements of this quantity have been undertaken during the
past few years, and the number and accuracy of available observations are still
rapidly increasing.

Photometric redshift estimates based on Lyman-limit systems (\cite{st96a,st96b})
stimulated the study of the rest frame UV continuum flux around 2000 {\AA} as a
preferred wavelength range for objects in the far universe.  This emission traces
the presence of massive (and therefore young) stars and can be related directly to
the actual star formation rate.  It is relatively easy to detect at high $z$ as it
is redshifted into the optical band.  However, older stellar populations and AGN
can make a significant contribution as well, leading to an overestimation of the
total SFR.  On the other hand, dust enshrouding star formation regions absorb very
efficiently the UV light, re-emitting it in the far IR.  Unfortunately, there is
still some uncertainty to account for dust extinction and its evolution with
redshift (\cite{nag01}).

Recombination lines from \ion{H}{II} regions can be in principle a more reliable
estimator of the instantaneous SFR, since the ionizing radiation
($\lambda\leq912${\AA}) comes from more massive (younger) stars than the softer UV
continuum.  These optical lines are less dramatically affected by dust extinction
and can be mapped with higher resolution in the local universe, but they require
infrared spectroscopy at moderate redshifts, which has not been possible until
very recently due to the faintness of the sources involved (\cite{pet98}).  For
H$_{\rm \alpha}$, this happens at $z\sim0.5$, and for forbidden lines, such as
[\ion{O}{II}]3727{\AA}, at $z\sim1.5$.  There is also a considerable degree of
uncertainty in the relation between recombination lines and star formation
activity.

Finally, the far infrared (FIR) spectrum of a galaxy can be roughly separated into
two components, namely the thermal emission of dust (heated by star formation
bursts) around $\lambda\sim60\ \mu$m and an infrared cirrus powered by the global
radiation field, which dominates for $\lambda\geq100\ \mu$m.  Neglecting cirrus
and AGN contamination, the FIR luminosity would also be an excellent tracer, but
instrumentation is not as developed in the IR and radio bands as it is in the
optical, and the conversion factor between FIR luminosity and SFR is for the
moment rather model dependent.

Although there is a large scatter among different indicators at any given
redshift, basically all studies find a significant increase (by about one decade)
in the cosmic SFR density from the present day to $z=1$.  However, it is still a
controversial issue (\cite{hop01}) whether it reaches a broad maximum there
(\cite{mad96}; \cite{gis00}) or it declines gradually towards higher redshifts
(\cite{saw97}; \cite{pas98}; \cite{ste99}).

From a theoretical point of view, star formation is a highly non-linear process,
which precludes any simple analytical treatment.  Most efforts towards modeling
the star formation history of the universe resort to semi-analytical methods or
numerical simulations to tackle the feedback mechanisms that self-regulate the SFR
within the hierarchical clustering scenario of structure formation.

Semi-analytical models (\cite{kau93}; \cite{col94}) construct a large sample of
Monte Carlo realizations of halo merging histories, using the Press-Schechter
formalism.  Hydrodynamics, star formation and feedback are implemented through
simple recipes, whose parameters are fixed in order to match the observed
properties of real galaxies, specially the luminosity function and the
Tully-Fisher relation (\cite{som99}).  Their high computational efficiency allows
a fast exploration of the parameter space in search of a viable model, but the
number of free parameters asks for supplementary modeling based on more
fundamental physics.

Numerical simulations are intended to solve directly the set of differential
equations of gravitation and gas hydrodynamics, and hence require much fewer model
assumptions than the semi-analytic approach.  However, despite the staggering
advances in computer technology and numerical algorithms, resolution in
large-scale hydrodynamical simulations is severely restricted by computational and
data storage requirements.  Although gas can be treated self-consistently, some
phenomenological recipes are still required in order to model star formation and
feedback.

The cosmic SFR density is a direct outcome of semi-analytical methods
(\cite{som01}) and hydrodynamical simulations (\cite{nag01}).  Both of them
predict a significant star formation activity at high $z$, favoring the hypothesis
of a gradual evolution of the comoving SFR as a function of time.

The present analysis is based on a series of numerical simulations, within
different cosmological models.  In previous papers, we have studied basic
observable properties of simulated galaxy samples, such as the luminosity function
and color dependence on environment (\cite{el99a}), and the Tully-Fisher relation
(\cite{el99b}).  In this paper, we will focus on the evolution of the SFR on a
cosmological scale, comparing our results with recent observations.

The following section is devoted to the observational estimation of the cosmic
SFR.  The standard procedure is briefly summarized and a compilation of recent
results is included with an analytical prescription for the transformation between
cosmological models.  Sect.~\ref{sims} describes the code and the different
numerical experiments on which the present paper is based.  Our results are
presented in Sect.~\ref{results}, where we compare with observations and
evaluate the separate contribution of cosmology, feedback effects and environment.
Our main conclusions are summarized in Sect.~\ref{conclus}.


\section{Observations}
\label{obs}

The first step to obtain an estimate of the SFR density from observations is the
selection of a {\em complete} sample of galaxies at a given redshift and
wavelength.  Some authors (\cite{ste99}) claim that the HDF covers a very small
area ($\sim5$ arcmin$^2$), and hence it is not statistically representative of the
whole universe.  This cosmic variance is difficult to quantify, but the strong
clustering of high redshift galaxies (and the fact that most estimates beyond
$z=2$ are based on the HDF) seems to indicate that it could be an important
effect.

Another crucial issue is the correction for incompleteness in the construction of
the luminosity function, which can be done following the $V_{\rm max}$ formalism
(\cite{sch68}) or resorting to more elaborate Monte Carlo algorithms (e.g.
\cite{ste99}).  The comoving luminosity density is easily obtained by integration
over all magnitudes (extrapolating the faint end as a Schechter function), and
then it can be converted to a SFR density with the help of a population synthesis
model.

A compilation of recent results is given in Table \ref{tableSFR}.  The first
column indicates the bibliographic reference, as well as the survey from which the
galaxy sample was extracted.  Next columns correspond to the chosen wavelength and
redshifts of each analysis, and finally the comoving luminosity and SFR densities.

\begin{table*}
\begin{center}
\begin{tabular}{lccccc}
 {\sc ~~ Survey} & {\sc Estimator}
 & $z$ & $\log{\rho_{\rm L}^{\rm SCDM}}$ &
$\log{\dot\rho_*^{\rm SCDM}}$ & $\log{\dot\rho_*^{\rm \Lambda CDM}}$ \\
\hline\\
HDF (\cite{pas98}) $^{\rm d}$ & 1500 {\AA}
  & $0.25 \pm 0.25$ & $26.47^{+0.31}_{-0.22}$ & -1.38 & -1.36\\
& & $0.75 \pm 0.25$ & $26.61^{+0.23}_{-0.14}$ & -1.24 & -1.26 \\
& & $1.25 \pm 0.25$ & $26.80^{+0.24}_{-0.12}$ & -1.05 & -1.10 \\
& & $1.75 \pm 0.25$ & $26.83^{+0.24}_{-0.12}$ & -1.02 & -1.08 \\
& & $2.50 \pm 0.50$ & $26.62^{+0.28}_{-0.21}$ & -1.23 & -1.30 \\
& & $3.50 \pm 0.50$ & $26.59^{+0.34}_{-0.27}$ & -1.26 & -1.34 \\
& & $4.50 \pm 0.50$ & $26.70^{+0.44}_{-0.37}$ & -1.15 & -1.23 \\
& & $5.50 \pm 0.50$ & $26.34^{+0.59}_{-0.38}$ & -1.51 & -1.59 \\
HDF (\cite{mad98}) $^{\rm d}$ & 1500 {\AA} & $2.75\pm0.75$ 
 & $26.90\pm0.15$ & -0.95 & -1.02 \\
& & $4.00\pm0.50$ & $26.50\pm0.20$ & -1.35 & -1.43\\
(\cite{ste99}) & 1700 {\AA} & $3.04\pm0.35$ & $27.04\pm0.07$ & -0.81 & -0.89\\
& & $4.13\pm0.35$ & $26.95\pm0.10$ & -0.90 & -0.98\\
(\cite{tre98}) $^{\rm d}$ & 2000 {\AA} 
& $0.15\pm0.15$ & $26.45\pm0.15$ & -1.40 & -1.35 \\
CFRS (\cite{lil96}) $^{\rm d}$ & 2800 {\AA} & $0.35 \pm0.15$ 
 & $26.14\pm0.07$ & -1.71 & -1.68\\
& & $0.625\pm0.125$& $26.46\pm0.08$ & -1.39 & -1.39\\
& & $0.875\pm0.125$& $26.78\pm0.15$ & -1.07 & -1.10\\
HDF (\cite{cow99}) $^{\rm d}$ & 2800 {\AA} & $0.35 \pm0.15$
 & $25.91\pm0.12$ & -1.69 & -1.66\\
& & $0.625\pm0.125$& $26.15\pm0.12$ & -1.45 & -1.45\\
& & $0.875\pm0.125$& $26.06\pm0.12$ & -1.54 & -1.57\\
& & $1.25\pm0.25$ &  $26.35\pm0.10$ & -1.25 & -1.30\\
HDF (\cite{con97}) $^{\rm d}$ & 2800 {\AA} & $0.75\pm0.25$ 
 & $26.77\pm0.15$ & -1.08 & -1.10 \\
& & $1.25\pm0.25$ & $26.94\pm0.15$ & -0.91 & -0.96\\
& & $1.75\pm0.25$ & $26.84\pm0.15$ & -1.26 & -1.32\\
HDF (\cite{saw97}) $^{\rm d}$ & 3000 {\AA} & $0.35\pm0.15$ 
 & $26.51\pm0.47$ & -1.34 & -1.31 \\
& & $0.75\pm0.25$ & $26.74\pm0.06$ & -1.11 & -1.13\\
& & $1.50\pm0.50$ & $26.93\pm0.05$ & -0.92 & -0.97\\
& & $2.50\pm0.50$ & $27.28\pm0.06$ & -0.57 & -0.64\\
& & $3.50\pm0.50$ & $26.91\pm0.10$ & -0.94 & -1.02\\
(\cite{gal95}) & H$_{\rm \alpha}$ & $0.025\pm0.025$&$39.09\pm0.04$ & -2.01 & -1.89 \\
CFRS (\cite{trm98}) & H$_{\rm \alpha}$ 
 & $0.2\pm0.1$ & $39.44\pm0.04$ & -1.66 & -1.60 \\
CFRS (\cite{gla99}) & H$_{\rm \alpha}$ 
 & $0.875\pm0.125$& $40.01\pm0.15$ & -0.91 & -0.94\\
(\cite{yan99}) & H$_{\rm \alpha}$ & $1.3\pm0.5$ & $40.21\pm0.13$ & -0.89 & -0.94 \\
CFRS (\cite{ham97}) $^{\rm d}$ & \ion{O}{II} 3727 {\AA}
 & $0.35 \pm 0.15$ & $38.63^{+0.06}_{-0.08}$ & -1.97 & -1.94 \\
& & $0.625\pm 0.125$ & $39.16^{+0.11}_{-0.15}$ & -1.44 & -1.44\\
& & $0.875\pm 0.125$ & $39.56^{+0.20}_{-0.38}$ & -1.04 & -1.07\\
CFRS (\cite{flo99}) & 15 $\mu$m 
 & $0.35\pm0.15$ & $41.97\pm0.25$ & -1.46 & -1.43\\
& & $0.625\pm0.125$& $42.20\pm0.25$ & -1.15 & -1.15\\
& & $0.875\pm0.125$& $42.53\pm0.25$ & -0.82 & -0.85\\
HDF (\cite{hug98}) & 60 $\mu$m & $3.0 \pm 1.0$
 & $\leq 42.26$ & -0.98 & -1.05 \\
(\cite{haa00}) & 1.4 GHz & $0.28\pm0.12$
 & $26.75^{+0.14}_{-0.21}$ & -1.17 & -1.13\\
& & $0.46\pm0.05$ & $27.03^{+0.24}_{-0.21}$ & -0.89 & -0.87\\
& & $0.60\pm0.05$ & $27.12^{+0.16}_{-0.26}$ & -0.80 & -0.80\\
& & $0.81\pm0.08$ & $27.39^{+0.13}_{-0.18}$ & -0.53 & -0.55\\
& & $1.60\pm0.64$ & $27.54^{+0.13}_{-0.18}$ & -0.38 & -0.44\\
\end{tabular}
\end{center}
\caption{Observational estimates of the cosmic SFR density at different
epochs. $\rho_{\rm L}$ refers to the luminosity density at the appropriate
wavelength, expressed in erg s$^{-1}$Hz$^{-1}$ Mpc$^{-3}$. $\dot\rho_*$
represents the comoving SFR density (in M$_\odot$ yr$^{-1}$ Mpc$^{-3}$)
for our SCDM and $\Lambda$CDM cosmologies. $^{\rm d}$ Original data has been
corrected for dust extinction.}
\label{tableSFR}
\end{table*}

Conversion factors between the last two quantities differ significantly from
author to author, and in some cases only the luminosity density was provided.  If
there is a SFR estimate in the original paper, then it is quoted in Table
\ref{tableSFR} as {\em corrected by dust extinction}; else, we follow Kennicutt's
(1998) prescription for an exponential burst and a Salpeter IMF between 0.1 and
100 M$_\odot$:
\begin{equation}
 \dot\rho_*\ [\rm{M}_\odot\ yr^{-1}]=\cases{
 1.4\times10^{-28}L_{\rm UV}\ [\rm{erg\ s^{-1}Hz^{-1}}] \cr
 1.4\times10^{-41}L_{3727}\ [\rm{erg\ s^{-1}Hz^{-1}}]}
\end{equation}

For dust extinction, we have applied a correction of A(1500-2000{\AA})=1.2 mag and
A(2880{\AA},\ion{O}{II})=0.625 mag, which correspond to factors of 3.02 and 1.78
respectively.  All modifications to the original data are shown as a footnote on
the bibliographic reference in the Table \ref{tableSFR}.

Most observational values given in Table \ref{tableSFR} were calculated assuming a
standard cold dark matter (SCDM) cosmology with dimensionless Hubble parameter
$h=0.5$ and mean matter density $\Omega_{\rm m} = 1$.  Since the conversion factor from
luminosity to SFR and the correction for dust extinction are completely
independent of the cosmological scenario, the computation of the comoving SFR
density in a $\Lambda$CDM ($h=0.7$) model only involves the transformation of the
luminosity densities.

In a flat universe, the volume enclosed by a solid angle $\Delta\omega$ between
$z-\Delta z$ and $z+\Delta z$ is

\begin{equation}
 V(z,\Delta z)=\frac{4\pi}{3}\Delta\omega\left( d_{\rm m}^3(z+\Delta
z)-d_{\rm m}^3(z-\Delta z)\right),
\end{equation}
where $d_{\rm m}(z)$ is the comoving distance in each cosmological model, while
luminosity scales with luminosity distance as $d_{\rm L}(z)=(1+z)d_{\rm m}(z)$.  Then,
assuming that all galaxies are located at the centers of the interval in $z$, the
luminosity density (and hence the total SFR density) will be proportional to
\begin{equation}
 \dot\rho_*(z)\propto\frac{L(z)}{V(z,\Delta z)}\propto\frac{d_{\rm m}^2(z)}{d_{\rm m}^3(z+\Delta
z)-d_{\rm m}^3(z-\Delta z)}
\label{ecLambda}
\end{equation}

To first order in $\Delta z$, the conversion between SFR densities in $\Lambda$CDM
and variants of the SCDM ($h=0.5$) cosmologies can be approximated as
\begin{equation}
 \frac{\dot\rho_*^{\rm \Lambda CDM}}{\dot\rho_*^{\rm SCDM}}\sim\frac{h^{\rm \Lambda
 CDM}}{h^{\rm SCDM}}\ \sqrt{\Omega_{\rm m}+\Omega_\Lambda(1+z)^{-3}}
\label{ecVolker}
\end{equation}

\begin{figure}
\resizebox{\hsize}{!}{\includegraphics{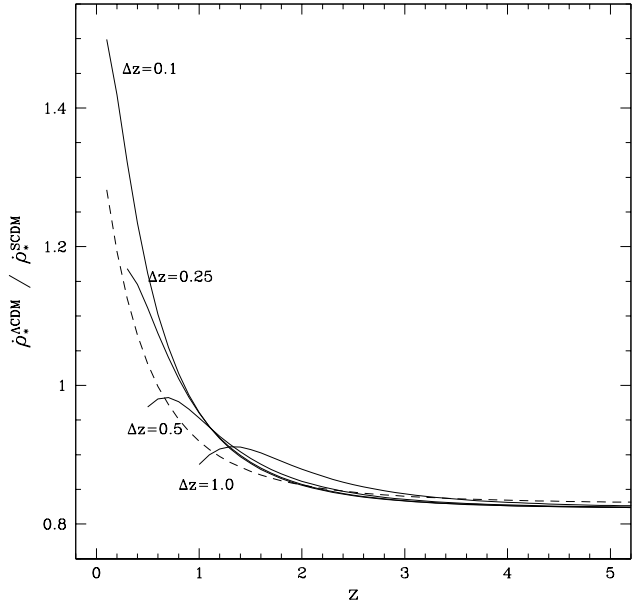}}
\caption{Conversion factor between star formation densities in our cosmological
models SCDM ($h=0.5$, $\Omega_{\rm m}=1$, $\Omega_\Lambda=0$) and $\Lambda$CDM 
($h=0.7$, $\Omega_{\rm m}=0.35$, $\Omega_\Lambda=0.65$) for several redshift intervals 
$\Delta z$. Dashed line corresponds to the approximation (\ref{ecVolker}).}
\label{volker}
\end{figure}

Fig.~\ref{volker} shows this factor, as well as the exact expression for different
values of $\Delta z$, which requires a numerical evaluation of the luminosity
distance in the $\Lambda$CDM model.  Although the results are very similar, we
have chosen to use the exact expression (\ref{ecLambda}) to compute the SFRs in
Table \ref{tableSFR} in order to minimize errors.

At low redshifts, the data are consistent for all the independent tracers of the
star formation activity once the effect of dust obscuration is taken into account.
The steep increase in the cosmic SFR density from the present day to $z\sim1$ is
firmly established.  However, there is not yet a full agreement about whether
there is a characteristic epoch of star formation in the universe around
$z\sim1.5$ or on the contrary that the SFR declines very slowly towards higher
$z$, as more recent observations seem to point out.  The results of ongoing deep
surveys will be extremely helpful in order to reach a definitive conclusion from
the observational point of view.


\section{Simulations}
\label{sims}

\subsection{Cosmological Models}

Up to now, the $\Lambda$CDM model has proven to be very successful in describing
most of the observational data at both low and high redshifts.  Moreover, recent
observations indicate the existence of a positive cosmological constant
(\cite{per99}; \cite{mel00}).  For our study we have chosen a spatially flat
universe with a cosmological constant $\Lambda/3H_0^2\equiv\Omega_\Lambda= 0.65$
and a present-day Hubble constant of $H_0=100h=70$~km/s/Mpc.  In order to study
the influence of cosmology on the global SFR, we have investigated three other
scenarios:  $\Lambda$CDM with lower $\Omega_{\rm \Lambda}= 0.6$, the standard CDM
model (dominated by dark matter density), and the BSI model with a {\em Broken
Scale Invariant} perturbation spectrum (as predicted by double inflation,
\cite{got91}).  The latter model is also dark matter-dominated, but it describes a
primordial power spectrum in which fluctuations on small scales are significantly
suppressed.  It is almost the same spectrum as in the $\tau$CDM model, in which
the primordial perturbation spectrum has been changed due to the hypothetical
decay of massive $\tau$ neutrinos (\cite{efs92}).  The model parameters are
summarized in Table \ref{tableSims}.  All spectra are COBE-normalized.  The
normalization is given by the linear dark matter mass variance $\sigma_8$ in a $8
h^{-1}$ Mpc sphere extrapolated to present time.

\begin{table*}
\begin{center}
\begin{tabular}{ccccccccccccc}
{\sc Experiment} & $n_{\rm r}$ & $\Omega_{\rm DM}$ &$\Omega_{\rm b}$ & $\Omega_\Lambda$ & $h$ &
$\sigma_8$ & $N$ & $L_{\rm box}$ & $L_{\rm cell}$ & $M_{\rm DM}$ & $A$ & ${\mathcal D}$ \\ 
\hline \\
$\Lambda$CDM1 & 1 & 0.325 & 0.025 & 0.65 
 & 0.7 & 0.9 & $350^3$ & 30 & 85.7 & 28.1 & 200 & 100\\
$\Lambda$CDM2 & 1 & 0.325 & 0.025 & 0.65 
 & 0.7 & 0.9 & $270^3$ &21.4& 79.3 & 22.2 & 200 & 100 \\
$\Lambda$CDM3 & 1 & 0.325 & 0.025 & 0.65 
 & 0.7 & 0.9 & $300^3$ & 12 & 40 & 2.85 & 200 & 100 \\
$\Lambda$CDM4 & 9 & 0.37 & 0.03 & 0.6 & 0.6 & 0.9 
 & $128^3$ & 5 &  39.1 & 2.22 & 200 & 100,500,1000 \\
$\Lambda$CDM5 & 18 & 0.324 & 0.026 & 0.65 & 0.7 
 & 0.9 & $128^3$ & 5 & 39.1 & 2.65 & 0,50,200 & 100 \\
SCDM & 18 & 0.95 & 0.05 & 0 & 0.5 & 1.2 
 & $128^3$ & 5 & 39.1 & 3.96 & 0,50,200 & 100 \\
BSI & 18 & 0.95 & 0.05 & 0 & 0.5 & 0.6 
 & $128^3$ & 5 & 39.1 & 3.96 & 0,50,200 & 100 \\
\end{tabular}
\end{center}
\caption{Collection of cosmological models, simulation specifics and feedback
parameters. The box length $L_{\rm box}$ is given in $h^{-1}$ Mpc, the cell
length $L_{\rm cell}$ is given in $h^{-1}$ kpc and the mass of dark matter particles
is given in $10^6 {\rm M}_\odot$.}
\label{tableSims}
\end{table*}

\subsection{Numerical code}

The code used to performed the numerical experiments consists of a combination of
a Particle-Mesh N-body gravity solver and an Eulerian {\em Piecewise Parabolic
Method} to account for gasdynamics.  A detailed modeling of the most relevant
non-adiabatic baryonic processes (i.e.  multiphase ISM, cooling, photoionization,
star formation and supernovae feedbacks) acting at a sub-grid level is also
incorporated.  The code is described in detail elsewhere (\cite{yep97}).  Here we
briefly review its main features regarding star formation and feedback.

The ISM is modeled as a two phase medium:  hot gas ($T_{\rm h} >10^4$ K) and a cold
phase ($T_{\rm c}=10^4$K), which pretends to mimic the molecular clouds from which stars
are formed.

Cold clouds can be formed from the hot gas by radiative and Compton cooling or by
thermal instability.  Radiative cooling rates, $\Lambda_{\rm r}(T,Z)$, are taken from
\cite{sut93} for two metallicity cases:  either primordial composition or solar,
depending on the star formation history of each computational element.

Stars are created, from the cold phase, as collisionless particles at each
timestep.  Their mass is proportional to the density of cold gas in each cell:
\begin{equation}
\dot\rho_*=-\dot\rho_{\rm c}=(1-\beta)\frac{\rho_{\rm c}}{t_*}
\end{equation}
where $\beta$ is the fraction of massive stars ($M >10M_\odot$) that explode 
as supernovae. For a Salpeter IMF, $\beta = 0.12$. For the characteristic time 
of star formation $t_*$ we assume a constant value of $10^8$ yrs.

Several feedback mechanisms are included:  photoionisation by UV light, gas
heating and evaporation by supernova explosions.  Massive stars are assumed to
create a homogeneously distributed background of ionizing radiation, and gas
should be dense enough to screen it in order to be able to cool into the cold
phase and form stars (e.g.  \cite{giroux}; \cite{muc97}).  A simple
prescription, without resorting to a detailed computation of photoionization
heating is to impose an overdensity threshold ${\mathcal D}$ for star formation 
to take place. On the other hand, supernova explosions can evaporate the cold 
gas phase and increase the temperature of the hot component.  In our model, the 
{\em supernova feedback parameter} $A$ controls the amount of energy (relative 
to the supernovae mass) that goes into cold cloud evaporation:
\begin{equation}
\left(\dot\rho_{\rm h}\right)_{\rm SN} = -\left(\dot\rho_{\rm c}\right)_{\rm SN} = 
  \frac{\beta\rho_{\rm c}}{t_*}A
\end{equation}
\begin{equation}
\left(\dot E_{\rm h} \right)_{\rm SN} = 
  \frac{\beta\rho_{\rm c}}{t_*}\left(\epsilon_{\rm SN}+A\epsilon_{\rm c}\right)
\end{equation}

where the subscripts $h$ and $c$ denote hot and cold gas, $\epsilon_{\rm SN} \simeq 
4.45\times10^{49}$ erg M$_\odot^{-1}$ is the supernovae specific energy 
(Salpeter IMF) and $\epsilon_{\rm c}$ is the specific energy of the cold gas phase 
($T_{\rm c}=10^4$ K).

\subsection{Numerical experiments}

Table \ref{tableSims} summarizes our set of experiments.  In order to improve
statistics, several realizations were run when the number of particles was low
enough ($N=128^3$), changing the random seeds used to generate the initial
conditions.  The number of realizations is given by $n_{\rm r}$ in the second column.
The cosmological model and the normalization $\sigma_8$ are shown in the next 5
columns, as well as the number of particles, box size, cell size (which determines
our spatial resolution) and mass of each dark matter particle (mass resolution).
In the last two columns we present the values of the supernova feedback parameter
$A$ and the overdensity threshold ${\mathcal D}$ required for star formation.

The comparison of experiments $\Lambda$CDM5, SCDM and BSI allows us to determine
the effect of cosmology and feedback from supernovae on the global SFR history.
Effects of photoionization are discussed from the results of experiment
$\Lambda$CDM4, and the dependence on resolution and volume can be studied from the
whole set of $\Lambda$CDM experiments.  In $\Lambda$CDM 1 and 2, the volume is
high enough to include a cluster of galaxies, so we have also investigated the
behavior of the SFR in different environments.

\subsection{Resolution effects}

In pure $N$-body simulations the lack of mass and force resolution leads to the
well known overmerging problem (see e.  g.  \cite{kly99}; \cite{kam01}), i.  e.
substructures of large halos or halos inside of halos are erased.

This effect becomes important in galaxy clusters.  To avoid overmerging high force
and mass resolution are necessary which may be reached later in adaptive mesh and
multimass N-body/gasdynamical codes.

Our code is based on a fixed Eulerian mesh, so resolution is limited to the cell
size.  High spatial resolution can only be achieved by increasing the number of
cells uniformly, or reducing the computational volume.  As a trade off between the
two, we have decided to restrict us on medium spatial resolution, in favour of
large simulation volumes.  Resolution effects are discussed in detail in the next
section.

The numerical code has been parallelized using OpenMP compiler directives.
Therefore, it runs very efficiently in parallel computers with either Shared or
Non Uniform Memory Access (NUMA) architectures.  The simulations have been
performed on a variety of machines.


\section{Results}
\label{results}

\subsection{Cosmology dependence}

\begin{figure*}
 \centering \includegraphics[width=14cm]{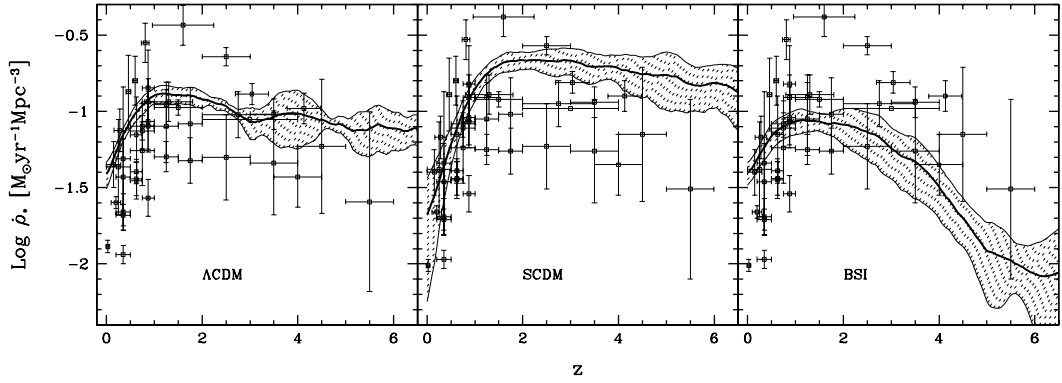}
\caption{Evolution of the comoving SFR density in different cosmologies
(experiments $\Lambda$CDM5, SCDM and BSI). The average value
over different
realizations is plotted as a thick solid line, while the shaded
area
represents the $1\sigma$ deviation from the mean. Dots correspond
to the
observational data points listed in Table \ref{tableSFR}}
\label{figCosmo}
\end{figure*}

The global star formation histories from simulations of the three cosmological
scenarios considered are compared in Fig.~\ref{figCosmo}.  The initial power
spectrum normalization determines the evolution of structures, which represent a
crucial factor in determining the rate of conversion of gas to stars.

Cosmological scenarios with CDM alone cannot explain structure formation on both
small and very large scales.  However, we have included SCDM as a reference model
in our comparison.  Due to its large power at small scales, SCDM produces many
structures at high redshift.  Therefore, we get a flat star formation rate even
beyond $z\sim6$, with a plateau about one decade over the observations
(Fig.~\ref{figCosmo}, middle panel).  It does not seem very likely that this
difference can be explained by dust absorption.  In this model, too many baryons
are transformed into stars too early with respect to observations, which poses an
additional problem for the SCDM model.

On the contrary, the BSI model leads to a clear maximum in the SFR at $z\sim2$,
but the absolute value is too small and the decrease at higher redshifts is
clearly too steep (Fig.~\ref{figCosmo}, right).  This model has not enough power
at small scales for producing enough stars at high redshift.  As a side effect,
the cosmic star formation rate density remains too high at low redshifts, since a
large part of cold gas remains in high density regions triggering further star
formation activity.

A cosmological model with non-zero cosmological constant offers the best agreement
with observations (Fig.~\ref{figCosmo}, left).  Both the overall trend and the
height of the plateau fit reasonably well to the observational data points.  At
present, the $\Lambda$CDM model fits best all known observational data (e.g.
\cite{bah99}).  Our results confirm that this model can also reproduce the
observed star formation history of the universe when dust extinction is taken into
account, in which case a significant amount of star formation is expected to take
place at high redshift.

A further feature (common to all cosmologies) is the smooth redshift dependence of
the mean SFR and the small range of scatter.  This is in sharp contrast to the
evolution of individual baryonic clumps, or the behavior of different regions in a
single time-step of the simulation.  There we observe huge bursts of star
formation with an increase in activity by a factor of 10 and higher, as well as a
quiescent regime with a star formation rate lower than the average.

In these simulations, the first stars are created around $z\sim10$, which seems to
be not too unrealistic, but it may depend on resolution.  This was not a primary
issue of the present study.

\subsection{Feedback dependence}

\begin{figure}[b]
 \resizebox{\hsize}{!}{\includegraphics{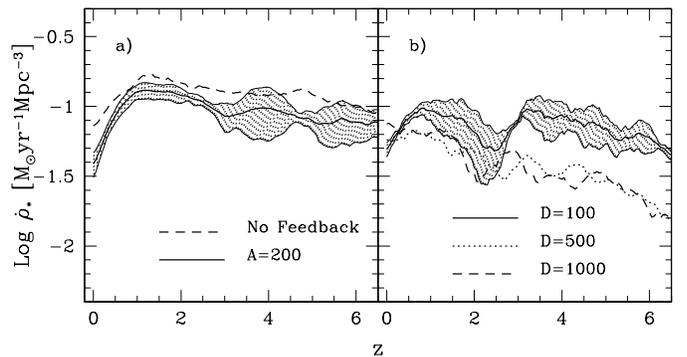}}
\caption{Effects of stellar feedback on the cosmic SFR density.
a) $\Lambda$CDM5 simulations with supernova feedback parameter
$A=200$ (solid
line) within $1\sigma $ error (shaded area) and $A=0$ (dashed
line).
b) Same as (a) but for $\Lambda$CDM4 simulations run with
different values
of the overdensity threshold $\mathcal D$ imposed by photoionization.}
\label{figFeedback}
\end{figure}

Fig.~\ref{figFeedback} shows the results of experiments $\Lambda$CDM5 and
$\Lambda$CDM4, from where the effects of supernova feedback and photoionization
were studied.  The same trends described in this section were also seen in SCDM
and BSI simulations.

Supernovae (left panel of Fig.~\ref{figFeedback}) play a key role in the slope of
the SFR density at low $z$.  Gas heating and evaporation act as a self-regulating
mechanism that inhibits subsequent gas infall when many new stars are formed.
This is specially important in the most massive objects, which show a much higher
star formation activity when feedback is not implemented.  Supernova explosions
can also blow away the gas reservoirs of dwarf galaxies.  The effect on the global
star formation rate is more noticeable at high redshift, when there are still very
few galaxies massive enough to retain the heated gas within their gravitational
potential wells.  However, photoionization (right panel) is the dominant effect at
very early epochs, preventing the gas from cooling and forming stars in low
density regions (\cite{efs2}).  As we have discussed above, our model takes
into account this effect through the overdensity threshold parameter ${\mathcal
D}$ for star formation.  Only in regions denser than ${\mathcal D}$, molecular
clouds will be able to form, and the onset of cosmic star formation activity is
thus retarded until more and more objects become dense enough to screen the
photoionizing background.

We found a SFR density consistent with observational data for ${\mathcal D}\sim
100$.  This is in roughly agreement with results from a more detailed theoretical
estimate of the same parameter (\cite{muc97}), although in that case the actual
value of ${\mathcal D}$, as well as the temperature range in which thermal
instability is an efficient process, depend on the local gas density and the
intensity of the ionizing flux.

\subsection{Environmental dependence}

In the hierarchical scenario of structure formation, the first objects collapsed
around the highest overdensity peaks and merged to eventually form the giant
ellipticals in the centers of clusters and the bulges of the most massive isolated
galaxies.  These objects contain the vast majority of stars older than $z=3$, and
hence their star formation history must be completely different from that of
galaxies living in less dense environments.

The experiments $\Lambda$CDM 1 and 2 sample a volume containing clusters,
filaments, and groups of galaxies as well as isolated galaxies.  {}\cite{got01}
have studied the merging history of dark matter halos as function of halo
environment.  They found that halos located inside clusters have formed earlier
than isolated halos of the same mass.  Moreover, they showed that at higher
redshifts ($z\sim 1-4$), progenitors of cluster and group galaxies have 3--5 times
higher merger rates than isolated galaxies.  Mergers of galaxies are thought to
play a crucial role in the evolution of galaxies.  In particular, the inflow of
material may serve as a source of fresh gas and therefore increase the star
formation rate.

\begin{figure}
\resizebox{\hsize}{!}{\includegraphics{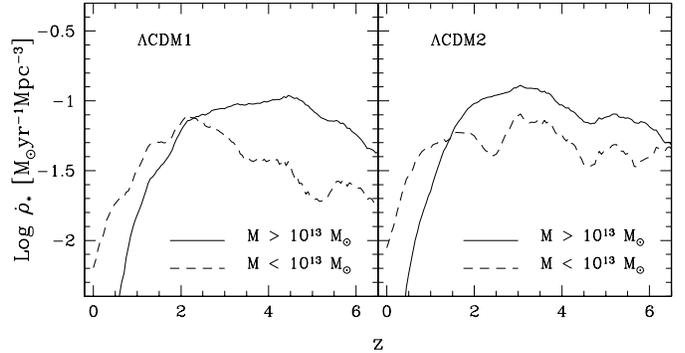}} 
\caption{Star formation rate densities in models $\Lambda$CDM1 and $\Lambda$CDM2
 due to halos more (solid lines) and less (dashed lines) massive than 
 $10^{13}$ M$_\odot$ at $z=0$.}
\label{figBIG} \end{figure}

In Fig.~\ref{figBIG} we plot the contribution to the cosmic SFR density of
progenitors of objects more massive than $10^{13}$ M$_\odot$ at present (clusters,
large groups) and less massive (isolated galaxies or small groups).  The most
massive objects have their peaks of star formation activity around $z\sim3$,
showing passive evolution since $z\sim2$.  This is in excellent agreement with the
merging history found by {}\cite{got01}.  The progenitors of these objects were
responsible for the bulk of star formation at high redshift, and can be very
likely associated with the Lyman Break Galaxy population.

Isolated galaxies and galaxies in small groups are younger.  They are responsible
for most of the current stellar production.  In these areas, halos massive enough
to retain their gas content have a nearly constant SFR (with sporadic burst
episodes).  

The simulated volume in our numerical experiments is too small to be a
representative cosmological volume.  In fact, the larger volume $\Lambda$CDM1
contains only one galaxy cluster of approximately $3\times 10^{14}$ M$_\odot$
whereas the samller volume $\Lambda$CDM2 contains four massive groups of about
$5\times 10^{13}$ M$_\odot$.  Comparing the solid lines in Fig.~\ref{figBIG} we
find qualitatively the expected scenario:  the peak of star formation in the
progenitors of the more massive cluster happened earlier (about $z = 4.5$)
than for the less massive groups ($z=3$).

In the absence of stellar feedback, most of the baryonic matter would have
cooled into small mini-galaxies (\cite{col91}; \cite{whi91}; \cite{bla92}).
Photoionization and supernova explosions inhibit cooling and star formation in
very low-mass objects, so early star formation was biased towards the high
overdensity regions that will collapse later to form the cores of the first
clusters.

As the collapse proceeds, gas is heated to temperatures that make the cooling
times too long and the SFR drastically drops once the cold gas has been
exhausted.

At the same time, new galaxies will form and produce stars.  Some of them
will remain in rather isolated areas until present (dashed lines in
Fig.~\ref{figBIG}) and some others will fall into the potential wells of clusters.
It seems very likely that tidal interactions trigger intense star formation bursts
in these infalling galaxies (\cite{lav88}), but ram pressure of the ICM may also
deprive them from their cold gas reservoirs (\cite{gun72}; \cite{qui00}).  This
later mechanism is supported by the observational evidence of the displacement of
the neutral hydrogen disk of post-starburst galaxies with respect to the stellar
population (\cite{val91}).

Cluster galaxies would then stop/decay their star formation activity and become
redder, giving rise to the \cite{but78} effect.  At the same time, they will
experience a morphological transformation, due to mergers with their neighboring
cluster galaxies, from spirals into S0 (\cite{dre97}).  At present, most of the
observed clusters are at $z\leq 0.5$ (\cite{fas00}, and references therein), so
deeper surveys are clearly needed in order to check this scenario.

\subsection{Resolution and volume dependence}

Star formation and feedback processes are incorporated in gasdynamical simulations
(either SPH or Eulerian) by means of rather simple recipes which pretend to {\em
extrapolate} the effects of local physics acting on scales of a few pc and masses
of $\sim 10^5$ M$_\odot $ below the resolution limit (often at kpc and higher).
In our case, part of this extrapolation is hidden in the two parameters:  The
supernova feedback parameter $A$ and the overdensity threshold for star formation
${\mathcal D}$.  Whether this parameterization is a reasonable extrapolation of
the underlying subgrid physics depends on the stability of results against changes
in spatial resolution of the simulations.

\begin{figure}
\resizebox{\hsize}{!}{\includegraphics{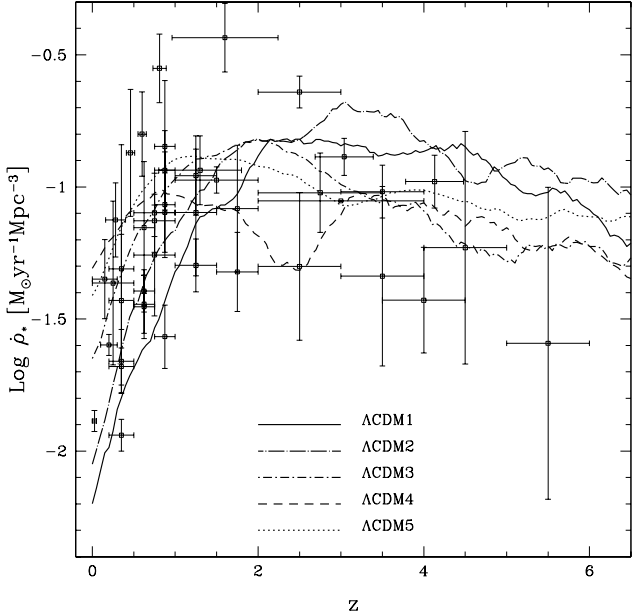}}
\caption{Comoving SFR in our 5 $\Lambda$CDM experiments, all with $A=200$ and
${\mathcal D}=100$. A significant star formation activity is found at high
$z$ regardless of resolution and volume. This is consistent with the full set
of observational data points (Table \ref{tableSFR}).}
\label{figCSFR}
\end{figure}

On the other hand, when one tries to estimate cosmological averaged quantities,
like the SFR density, it is necessary to check for deviations of these quantities
due to small number statistics.  In the case that matters here, a volume averaged
quantity, the {\em cosmic variance} can have a relatively large contribution in
the particular determination of the comoving SFR density.  Thus, we have simulated
different computational volumes with similar resolutions, depending on the
computational facilities.

To check for these two effects we compare in Fig.~\ref{figCSFR} our results for
$\Lambda$CDM simulations with different resolutions and volumes.

The most noticeable feature that we derive from this Figure is the steeper slope
of the local SFR for larger volume simulations.  As we increase the volume, larger
structures are being simulated (i.e.  clusters) and as explained above, the
reduced star formation activity in the cores of clusters in recent epochs can
explain this behavior.  In simulations with smaller volumes, most of the galaxies
that form are either isolated or in small groups, which have a completely
different star formation activity.  Thus, we can conclude from our simulations
that the effects of cosmic variance are quite important in determining the proper
behavior of the cosmic SFR density from $z\geq 1$ to the present.

We do not find significant differences in our estimates of the SFR density from
simulations with different volumes and spatial resolutions, but there is a general
trend to have more SFR density at early times for simulations with higher
resolution. This is a consequence of the overcooling problem (see
\cite{bal01} for a recent review), since the number of resolved low mass halos
increases in these simulations.  A detailed analysis of this question would
require a more realistic treatment of photoionization.

Keeping all these considerations in mind, we should be extremely cautious when
making {\em quantitative} estimates of the cosmic SFR density at any given $z$.
However, this quantity shows a similar behavior in all $\Lambda$CDM experiments.
This is a very promising result indicating that its {\em qualitative} evolution
(particularly at high redshift) is a robust prediction of our simulations.

No characteristic epoch of cosmic star formation is found for a $\Lambda$CDM
universe.  The comoving SFR density increases with lookback time until it reaches
a nearly constant mean plateau beyond $z=2$.  Stars probably formed gradually at
high redshift, according to most recent observations and taking dust extinction
into account.  At some point around $z\sim1$, the global star formation activity
would have declined sharply until the present day, although individual SFRs of
active galaxies are kept approximately constant.

The star density $\rho_*(z)$ shows a similar behavior for both CDM and
$\Lambda$CDM models.  The most remarkable difference is found in the present value
of $\rho_*(0)$ (see Table \ref{rhoSt}), which can be observationally deduced from
the local luminosity density, assuming a constant mass-to-light ratio
(\cite{mad96}; \cite{fuk98}):
\begin{equation}
 \rho_*^{\rm obs}(0)\sim7\times10^8\ h\ {\rm M}_\odot\ {\rm Mpc}^{-3}
\end{equation}

Based on the fraction of stars in bulges, ellipticals and S0s, \cite{ren98}
estimated that about $30\%$ of the stellar content of the universe should have
already been formed by $z=3$, regardless of the cosmological model.  The amount of
stars produced in experiments $\Lambda$CDM1 and 2 (Table \ref{rhoSt}), both at
$z=3$ and $z=0$, are in fair agreement with these observationally derived
quantities.  For the other experiments, the fraction of stars older than $z=3$ is
significantly lower, which is not only due to a low star formation activity at
high redshift, but also to a higher SFR in the local universe and the effects of
cosmic variance due to the smaller simulated volumes.

\begin{table}
\begin{center}
\begin{tabular}{ccc}
{\sc Experiment} & $\rho_*(z=0)$ & $\rho_*(z=3)/\rho_*(0)$ \\ \hline \\
 $\Lambda$CDM1 & 6.50 & 0.252 \\
 $\Lambda$CDM2 & 8.60 & 0.243 \\
 $\Lambda$CDM3 & 9.51 & 0.115 \\
 $\Lambda$CDM4 & 9.01 & 0.116 \\
 $\Lambda$CDM5 & 10.9 & 0.119 \\
 SCDM & 12.3 & 0.115 \\
 BSI & 7.37 & 0.041 \\
\end{tabular}
\end{center}
\caption{Present star density (in $10^8\ h$ M$_\odot$ Mpc$^{-3}$) and
 fraction of stars older than $z=3$ for the different numerical experiments.}
\label{rhoSt}
\end{table}


\section{Conclusions}
\label{conclus}

We have performed 66 numerical simulations, grouped in 7 different experiments, in
order to obtain a theoretical prediction of the cosmic star formation history.  We
compared our predictions with a number of published observational estimates.  We
summarize our results as follows:

\begin{enumerate}

\item $\Lambda$CDM shows the best agreement with observations.  Standard CDM tends
to overpredict the cosmic SFR density, while BSI fails to form enough stars at
high $z$ due to its reduced power at small scales.

\item Photoionization fixes the onset of star formation in the universe,
 supressing star formation in low-mass objects at high $z$. Supernovae feedbacks
 are very efficient to self-regulate the process of converting gas into stars,
 preventing the most massive halos to form too many stars at later epochs.

\item Star formation histories are very different for objects that end up in the
 cores of clusters and isolated galaxies.  Galaxies in cluster cores have much
 older stellar populations, and their activity is drastically reduced when they
 lose their cold gas reservoirs.

\item In contrast to the first observational estimates (\cite{mad96}), but in
 agreement with most recent data (\cite{ste99}), the $\Lambda$CDM model predicts
almost no drop in the cosmic SFR for $2 < z < 5$. Star formation seems to be a
gradual process with no characteristic epoch.

\item The comoving simulated volumes in our experiments are comparable to those
 covered by observations.  Thus, one should expect a non negligible error in the
 observational measurements due to cosmic variance associated with small volume
 statistics.  This is most important when one wants to determine the steep slope
 of the SFR density from $z\sim 1$ to the present time.  \end{enumerate}

Selection effects, dust extinction and the conversion from luminosities to SFRs
are the major sources of uncertainty in the observational estimates.  The small
volume sampled by the HDF may also introduce some statistical bias at high
redshift.

Uncertainty in our numerical experiments comes from spatial resolution and the
star formation and feedback prescriptions.  The use of Adaptive Mesh Refinement
Eulerian codes (AMR) will be useful in order to improve resolution in
statistically significant cosmological volumes.  Photoionization and chemical
enrichment have been given a phenomenological treatment in our code.  It would be
extremely interesting to self-consistently model both processes, computing the
intensity of the ionizing background at each time-step, as well as metal
production and advection.

\acknowledgements

This work has been partially supported by the SEUID (Spain) under project number
PB96-0029, by the {\em Acciones Integradas Hispano-Alemanas HA2000-0026} and by
Deutscher Akademischer Austauschdienst DAAD.

We thank the Centro Europeo de Paralelismo de Barcelona and the Centro de
Computaci{\'o}n Cient{\'\i}fica de la Universidad Aut{\'o}noma de Madrid for
allowing us to use their computational facilities in which most of the simulations
reported in this paper were performed.


\end{document}